\documentclass[11pt,a4paper]{article}
\usepackage{jinstpub}
\usepackage[utf8]{inputenc}
\usepackage{float}
\usepackage{lineno}

\title{The $^{3}$He BF$_{3}$ Giant Barrel (HeBGB) Neutron Detector}
\author[1]{K. Brandenburg
\note{Corresponding author.}}
\author[]{G. Hamad,}
\author[]{Z. Meisel,}
\author[]{C.~R. Brune,}
\author[]{D.~E. Carter,}
\author[]{T. Danley,}
\author[]{J. Derkin,}
\author[]{Y. Jones-Alberty,}
\author[]{B. Kenady,}
\author[]{T.~N. Massey,}
\author[]{S. Paneru,}
\author[]{M. Saxena,}
\author[]{D. Soltesz,}
\author[]{S.~K. Subedi,}
\author[]{J. Warren}

\affiliation[]{Institute of Nuclear \& Particle Physics, Department of Physics \& Astronomy, Ohio University,
Athens, Ohio 45701, USA}
\emailAdd{kb851615@ohio.edu, meisel@ohio.edu}
\abstract{$(\alpha,n)$ reactions play an important role in nuclear astrophysics and applications and are an important background source in neutrino and dark matter detectors. Measurements of total $(\alpha,n)$ cross sections employing direct neutron detection often have a considerable systematic uncertainty associated with the energy-dependent neutron detection efficiency and the unknown initial neutron energy distribution. The $^{3}{\rm He}\,{\rm BF}_{3}$ Giant Barrel (HeBGB) neutron detector was built at the Edwards Accelerator Laboratory at Ohio University to overcome this challenge. HeBGB offers a near-constant neutron detection efficiency of ($7.5\pm 1.2$) \% over the neutron energy range 0.01~MeV -- 9.00~MeV, removing a significant source of systematic uncertainty present in earlier $(\alpha,n)$ cross section measurements.}
\begin{document}
\maketitle

\section{Introduction}

$(\alpha,n)$ reactions are a common feature of thermonuclear environments with a neutron excess. Temperatures near or above tens of megakelvin (MK) are sufficient for $\alpha$-particles to overcome the coulomb barrier and, for the neutron-rich nuclides in these environments, the $Q$-value for the neutron-emitting exit channel is generally the most favorable. For extreme neutron excesses, the reverse reactions $(n,\alpha)$, which are related to the forward $(\alpha,n)$ reactions via detailed balance, are themselves dominant. Furthermore, $\alpha$-emission from actinides present in fissile material or as contaminants leads to an additional neutron flux from subsequent $(\alpha,n)$ reactions on otherwise inactive surrounding material, complicating neutron flux analyses and contributing to detector backgrounds.
As such, $(\alpha,n)$ reactions play an important role for a variety of topics, including energy generation and diagnostics from nuclear fission and nuclear fusion~\cite{Serp14,Cerj18}, nucleosynthesis in astrophysical environments~\cite{Brav12,Blis20}, dark matter and neutrino detector backgrounds~\cite{Apri13,Hari05,West22}, and forensic analyses and predictions associated with nuclear explosives~\cite{Runk10,Dola14}. Meanwhile, a large fraction of $(\alpha,n)$ cross sections for stable targets have not been measured and many of the measured reactions have been shown to disagree with theory predictions~\cite{Pere16}.

($\alpha$,n) cross sections can be measured using various techniques including both activation and direct neutron detection. In an activation measurement, target nuclei are bombarded with a monoenergetic beam for some time. Gamma rays from the resulting radioactive sample are measured to determine the cross section. This method, while precise, is not applicable to all reactions of interest due to measurement time constraints created by either too short or too long lived decay products, or decay radiation that is overwhelmed by prominent background radiation. The stacked foil technique has been used to measure the cross section at several energies simultaneously with a single particle beam; however, a large reaction energy uncertainty is introduced for the lowest center of mass energies included in the measurement~\cite{Hagi18}.

Direct neutron detection often involves neutron long counters. Neutron long counters~\cite[e.g.][]{Utsu17,Arno11,Laur14,Gome11,Tari17,Fala13,Pere10,Math12} offer high efficiency and many are designed to have near 4$\pi$ solid angle coverage. In the long counter scheme, neutrons from the reaction are emitted from the target location and are moderated typically through a low atomic number ($Z$) material before being captured in neutron-sensitive proportional counters that are embedded within the moderator. Due to the moderating material, the initial energy of a neutron is lost before detection. In many cases the initial neutron energy distribution is unknown, so a large variation in efficiency with respect to initial neutron energy can lead to large relative uncertainties~\cite{Banu19}. When this is not taken into consideration, systematic uncertainties can be introduced into the measurement result~\cite{Pete17,Mohr18}.

The $^{3}{\rm He}$ BF$_{3}$ Giant Barrel (HeBGB) at the Edwards Accelerator Laboratory at Ohio University is a neutron long counter optimized to measure ($\alpha$,n) cross sections of interest for nuclear astrophysics and applications. HeBGB has been optimized to provide near-constant neutron detection efficiency for the relevant neutron energies.
Reaction-based validation measurements and source measurements have been performed to confirm the simulated efficiency of the detector. The following section will discuss the design process of the detector in MCNP6~\cite{Goor12}, validation measurements are discussed in Section~\ref{section:Experimental Validations}, efficiency corrections in Section~\ref{section:Efficiency Corrections},
average initial neutron energy determination in Section~\ref{section:Average Initial Neutron Energy Determination}, and conclusions are given in Section~\ref{section:Conclusions}.

\section{Design of HeBGB}
\label{section:Design}

\begin{figure}[ht]
        \begin{center}
        \includegraphics[width=.9\columnwidth,angle=0]{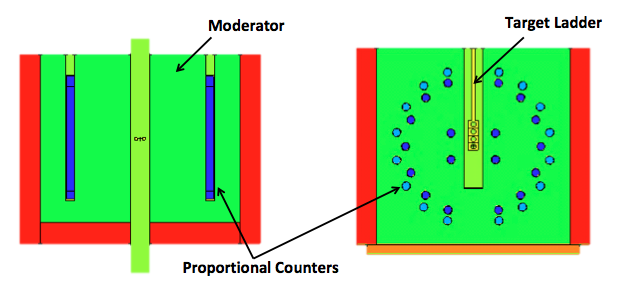}
        \caption{(color online) Cross sectional view of the HeBGB long counter. Left: top down view and right: view from the beam path. The moderating material shown in green is enclosed on 3 sides by borated polyethylene with thickness 3 cm and 5\% by weight boron loading shown in red. The bottom of the detector is held by a steel table shown in orange. BF$_{3}$ detectors make up the inner and middle rings shown as dark blue circles, while $^{3}$He detectors make up the outer ring shown as lighter blue circles in the right image. }
        \label{HeBGBView}
        \end{center}
        \end{figure}

The HeBGB moderator is made of four 5.0'' x 23.6'' x 23.6'' pieces of natural ultra-high-molecular-weight polyethylene having a density of (0.95 $\pm$ 0.2) g\,cm$^{-3}$. Embedded in the polyethylene are rings of 16 $^{3}$He and 18 BF$_{3}$ proportional counters and a T-style beam pipe which allows for passage of the beam and a target drive system as shown in Figure~\ref{HeBGBView}. The BF$_{3}$ proportional counters were previously used in Ref.~\cite{Mick03}. See Table~\ref{tab:i} for a full list of the detector specifications.

    \begin{table}[htbp]
    \centering
    \caption{\label{tab:i} Detector Specification for the $^{3}$He (model number SA - P4-0814-102 and operating voltage 975V) and BF$_{3}$ (model number RS-P1-0813-101 and operating voltage 2400V) proportional counters.}
    \smallskip
    \begin{tabular}{|lrrrrr|}
    \hline
    Type & Manufacturer & Pressure [atm] & Sensitive Length [in] & Length [in] & Diameter [in]\\
    \hline
    $^{3}$He & Baker Hughes & 4.00 & 14.00 & 16.19 & 1.00\\
    BF$_{3}$ & Reuter--Stokes & 0.723 & 12.25 & 14.44 & 1.00\\
    \hline
    \end{tabular}
    \end{table}

A large number of MCNP6 calculations were performed to optimize the configuration of the proportional counters within the moderator. The configurations were created by varying the number of rings, ring radii, and number of proportional counters in each ring. Exploratory calculations were performed for 2, 3 and 4 rings with ring radii anywhere between 5 to 28 cm, where the radial spacing between rings was a minimum of 3 cm to prevent overlap of the detectors within the simulation. The number of proportional counters in each ring varied between 2 and 16, with equal numbers in the left and right quadrants, and the location of the BF$_{3}$ and $^{3}$He counters in the inner versus outer rings was also explored. Based on these initial calculations, a smaller phase space with 3 rings was simulated in more detail with 0.1~cm to 0.5~cm steps in the ring radius.

For each configuration, 10,000 monoenergetic events were generated at each energy, from 10~keV to 9~MeV in 1~MeV steps, with an isotropic angular distribution. The neutron detection efficiency was evaluated at each energy and an average $\epsilon_{\rm avg}$ was evaluated over the energy range. The relative uncertainty in the efficiency $\delta\epsilon$ was defined as half of the spread between the maximum efficiency $\epsilon_{\rm max}$ and minimum efficiency $\epsilon_{\rm min}$ over the energy range:

    \begin{equation}
    \delta\epsilon = \frac{\epsilon_{\rm max} - \epsilon_{\rm min}}{2\epsilon_{\rm avg}}.
    \label{eqn:EfEr}
    \end{equation} 
    
We use $\epsilon_{\rm avg}$ and $\delta\epsilon$ as metrics to compare the anticipated performance of each proportional counter configuration.
 
\begin{figure}[ht]
        \begin{center}
        \includegraphics[width=.9\columnwidth,angle=0]{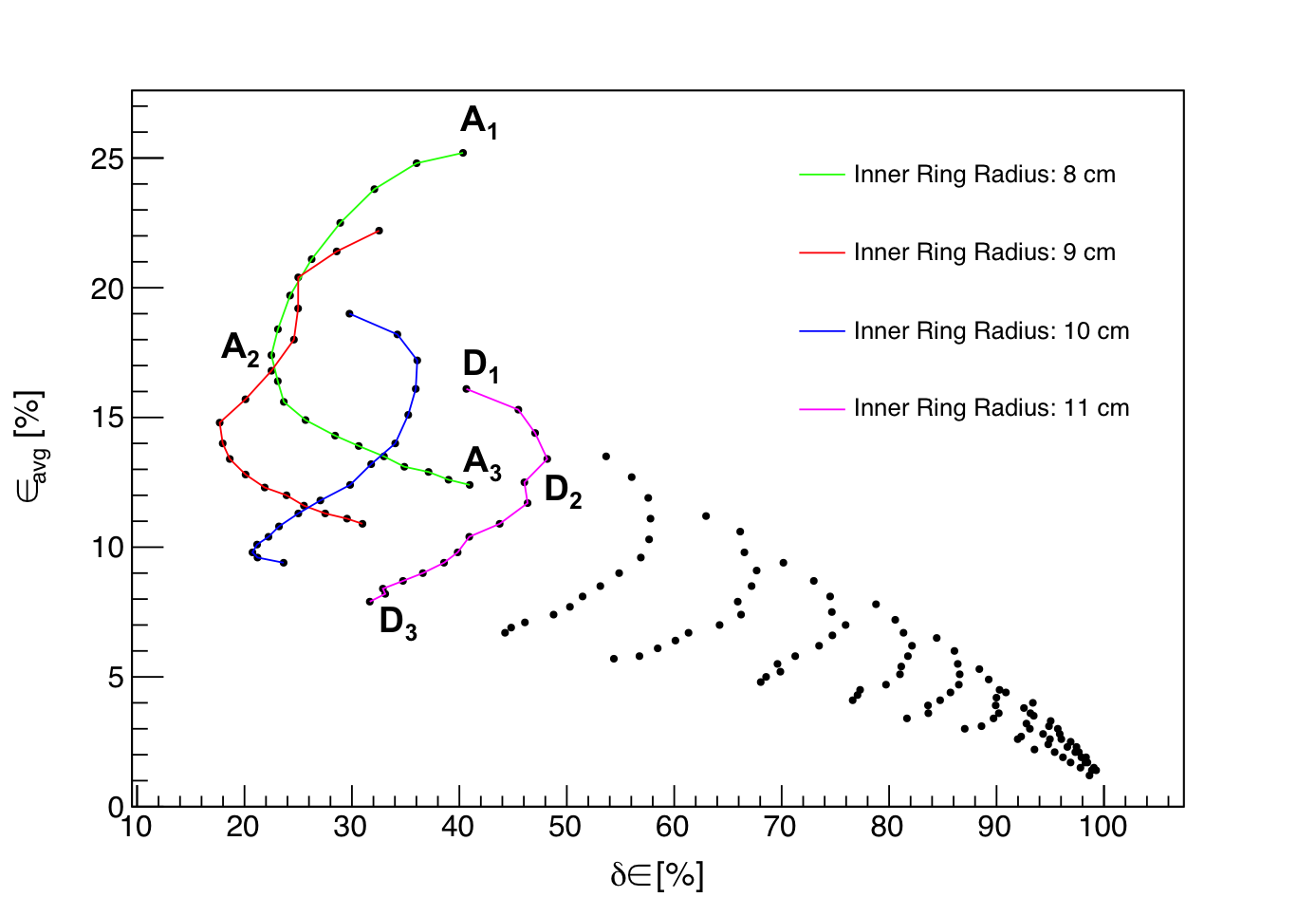}
        \caption{(color online) Sampled $\epsilon_{\rm avg}$ and $\delta\epsilon$ parameter space for the 2 ring configurations. Lines are drawn connecting the inner ring radius cases for 8, 9, 10, and 11 cm, each for various outer ring radii configurations. Sample configurations A$_{1-3}$ have outer ring radii 11, 18, and 28 cm, respectively, and sample configurations D$_{1-3}$ have have outer ring radii 14, 19, and 28 cm, respectively.
        \label{TwoRings}}
        \end{center}
        \end{figure}

 Before presenting the results of all calculations, we first consider an example subset, which involves one inner ring of BF$_{3}$ and one outer ring of $^{3}$He counters. These calculations were performed using two rings in each configuration, beginning with an inner ring radius of 8 cm and outer ring radius of 11 cm. For a fixed inner ring radius, the outer ring radius was increased in steps of 1~cm until the outer dimensions of the moderator are reached. The inner ring radius was then increased by 1~cm and the process repeated. Figure~\ref{TwoRings} shows $\epsilon_{\rm avg}$ and $\delta\epsilon$ for the example subset, where each point corresponds to a detector configuration and lines connect the results of calculations performed with the same inner ring radius.

Generally, smaller ring radii result in a larger $\epsilon_{\rm avg}$. This is because a) the density of detectors within a ring decreases within increasing ring radius and b) because of the neutron thermalization process. The majority of neutron-captures within a proportional counter occur for thermal neutron energies and the majority of neutrons are thermalized within a shorter path length. The average path length that a neutron of an initial energy travels before capture is the sum of the mean-free-paths $\ell_{\rm mfp}$ of the neutron over its scattering history. On average, for neutron energies below 10~MeV, each scattering event will reduce the neutron energy from its precollision energy by half~\cite{Knoll}. For a given energy $\ell_{\rm mfp}=1/\Sigma_{\rm s}$, where the macroscopic scattering cross section in polyethylene is
\begin{equation}
   \Sigma_{\rm s}=\frac{\rho N_{\rm A}}{M}(2\sigma_{\rm C}+4\sigma_{\rm H}).
    \label{eqn:ScatterCS}
\end{equation}

Here, $\rho=0.95$~g\,cm$^{-3}$ and $M=28$ g are the density and molecular weight of polyethylene and $\sigma_{\rm C}$ and $\sigma_{\rm H}$ are the microscopic neutron scattering cross sections from Ref.~{\cite{Chad11}}. Following this procedure, for a neutron with a 2~MeV initial energy, the average path length is 2.7~cm. For a 9~MeV neutron, the average path length is 7.5~cm. 

Meanwhile, $\delta\epsilon$ tends to decrease with increasing distance between the inner and outer ring radii. This follows from the preceding discussion of the average path length, where, for a fixed inner ring radius, increasing the outer ring radius will improve the efficiency for higher-energy neutron detection and thus make the overall response more uniform.

\begin{figure}[ht]
        \begin{center}
        \includegraphics[width=.9\columnwidth,angle=0]{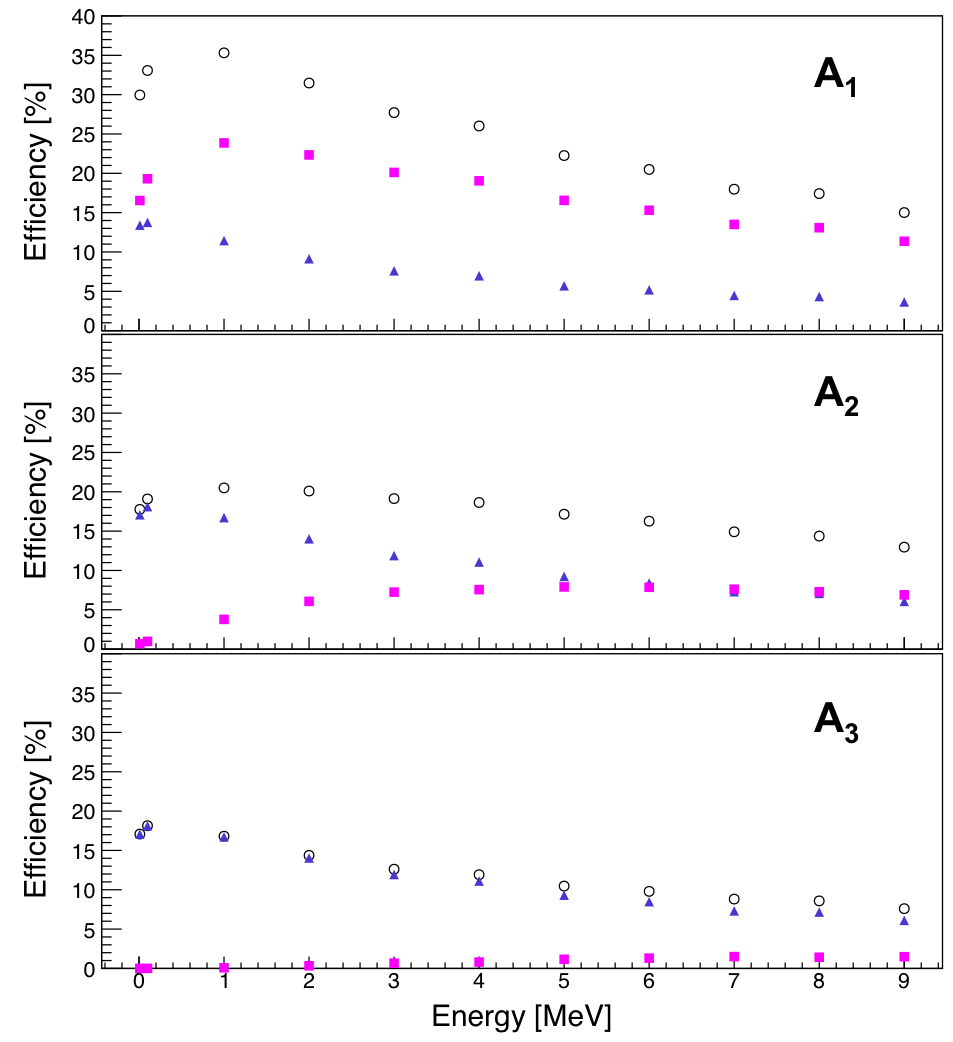}
        \caption{(color online) Efficiency curves for 3 sample cases along the green 8 cm inner ring radius line of Figure~\ref{TwoRings} . Open black circles are the total efficiency, blue triangles are the inner ring BF$_{3}$ efficiency and pink squares are the outer ring $^{3}$He efficiency having outer ring radii of 11, 18, and 28 cm for A$_{1-3}$, respectively.
        \label{fig:IR8}}
        \end{center}
        \end{figure}

\begin{figure}[ht]
        \begin{center}
        \includegraphics[width=.9\columnwidth,angle=0]{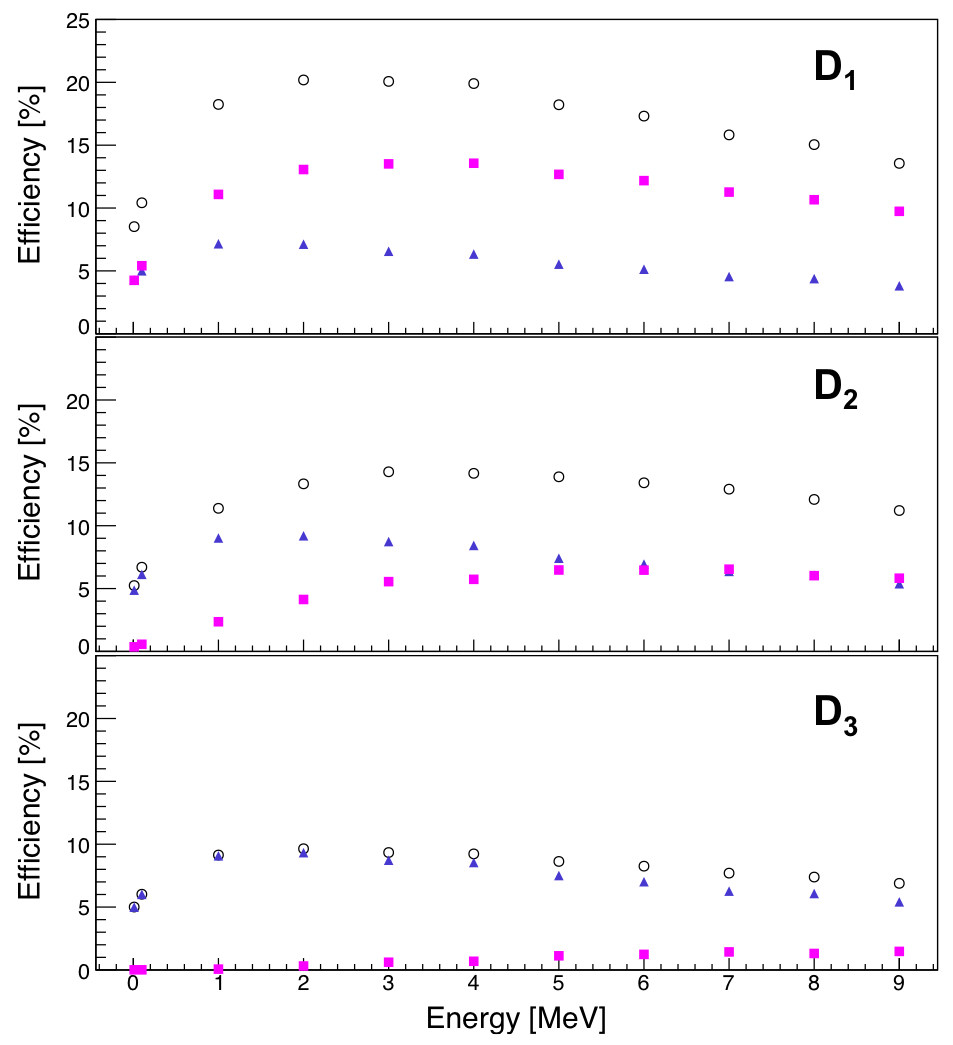}
        \caption{(color online) Efficiency curves for 3 sample cases along the pink 11 cm inner ring radius line of Figure~\ref{TwoRings}. Open black circles are the total efficiency, blue triangles are the inner ring BF$_{3}$ efficiency and pink square are the outer ring $^{3}$He efficiency having outer ring radii of 14, 19, and 28 cm for D$_{1-3}$, respectively.
        \label{fig:IR11}}
        \end{center}
        \end{figure}

The curvature within the $\epsilon_{\rm avg}$--$\delta\epsilon$ phase space for the subset of calculations featured in Figure~\ref{TwoRings} differs depending on the inner ring radius. Case A, with an 8~cm inner ring radius initially decreases in $\delta\epsilon$ with increasing outer ring radius, until finally increasing again. Case D, with an inner ring radius of 11~cm, has the opposite behavior. To understand this, we turn to the by-ring efficiencies shown in Figures~\ref{fig:IR8} and \ref{fig:IR11}. For case A, increasing the outer ring radius decreases $\epsilon$ at all energies, but the decrease is more rapid for low neutron energies. As such, $\delta\epsilon$ initially decreases. However, for the largest outer-ring radius, the outer ring efficiency drops significantly even for the highest energy neutrons, thus increasing $\delta\epsilon$. For case D, the inner ring alone already has a relatively flat $\delta\epsilon$, as the distance is within the path length distribution of some of the lowest and highest-energy neutrons. As such, it is advantageous for the outer ring radius to be large enough that it is only enhancing the efficiency for the highest energy neutrons.

\begin{figure}[ht]
        \begin{center}
        \includegraphics[width=.9\columnwidth,angle=0]{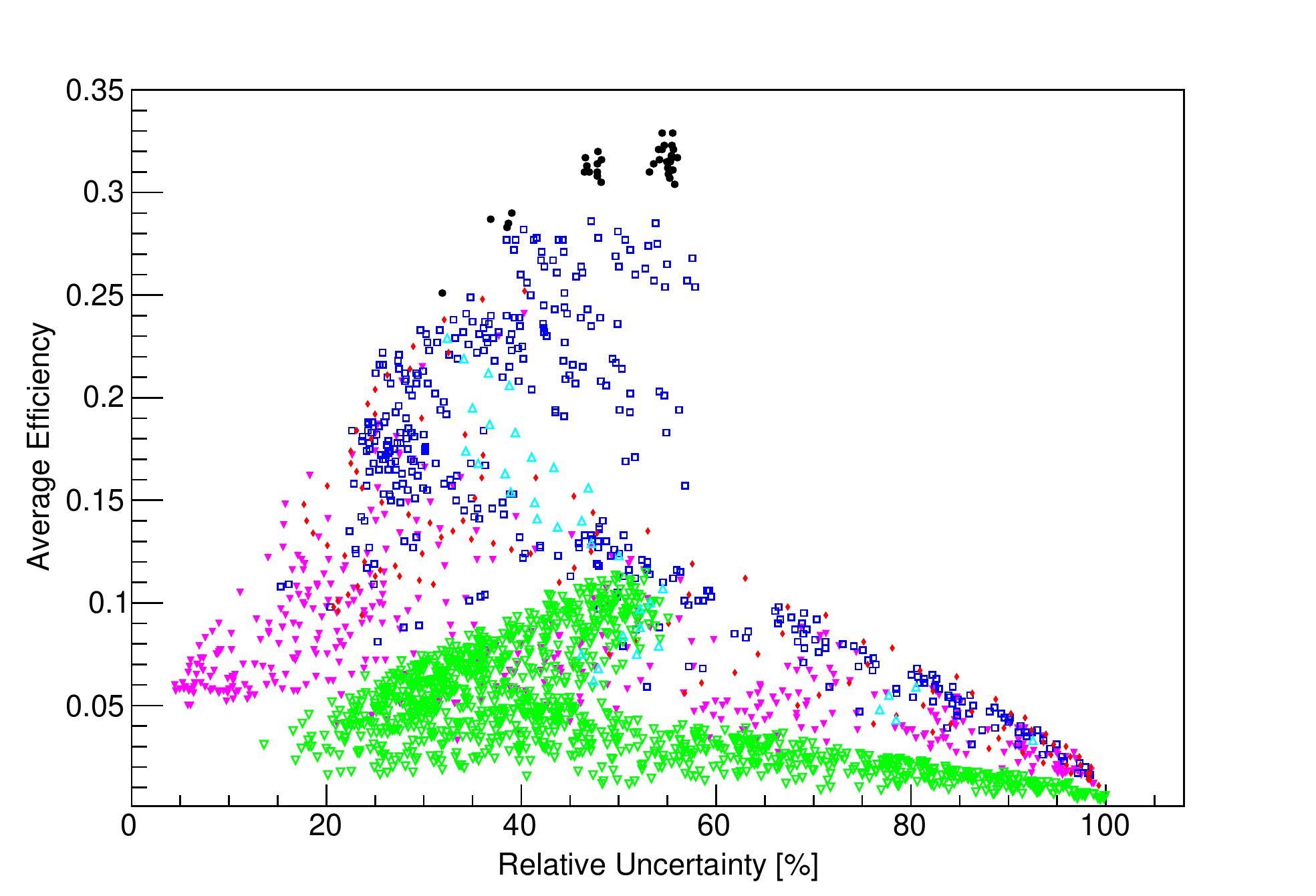}
        \caption{(color online) The average efficiency and relative uncertainty for each design configuration simulated. Black filled circles are 4 rings with equal numbers in each ring alternating between helium and boron. Cyan open up-pointing triangles are 4 rings with equal numbers in each ring with 2 inner rings of boron and 2 with helium . Green down-pointing open triangles are 3 rings with no helium counters. Blue open squares are 3 rings with 2 inner of helium and 1 outer of boron. Purple down-pointing filled triangles are 3 rings with 2 inner of boron and 1 outer of helium. Red filled diamonds are 2 rings with 1 inner of boron and 1 outer of helium. Note, the final design configuration contains 2 inner rings of boron and 1 outer ring of helium detectors. 
        \label{AllPoints}}
        \end{center}
        \end{figure}

Figure~\ref{AllPoints} shows the results for all configurations. Distributing the proportional counters amongst more rings generally enables a larger $\epsilon_{\rm avg}$, as detectors can be placed near the average path length for several neutron energies. However, increasing the number of rings ultimately forces inner rings to smaller radii, which increases $\delta\epsilon$ due to the high detection efficiency for low-energy neutrons.

\begin{figure}[ht]
        \begin{center}
        \includegraphics[width=.9\columnwidth,angle=0]{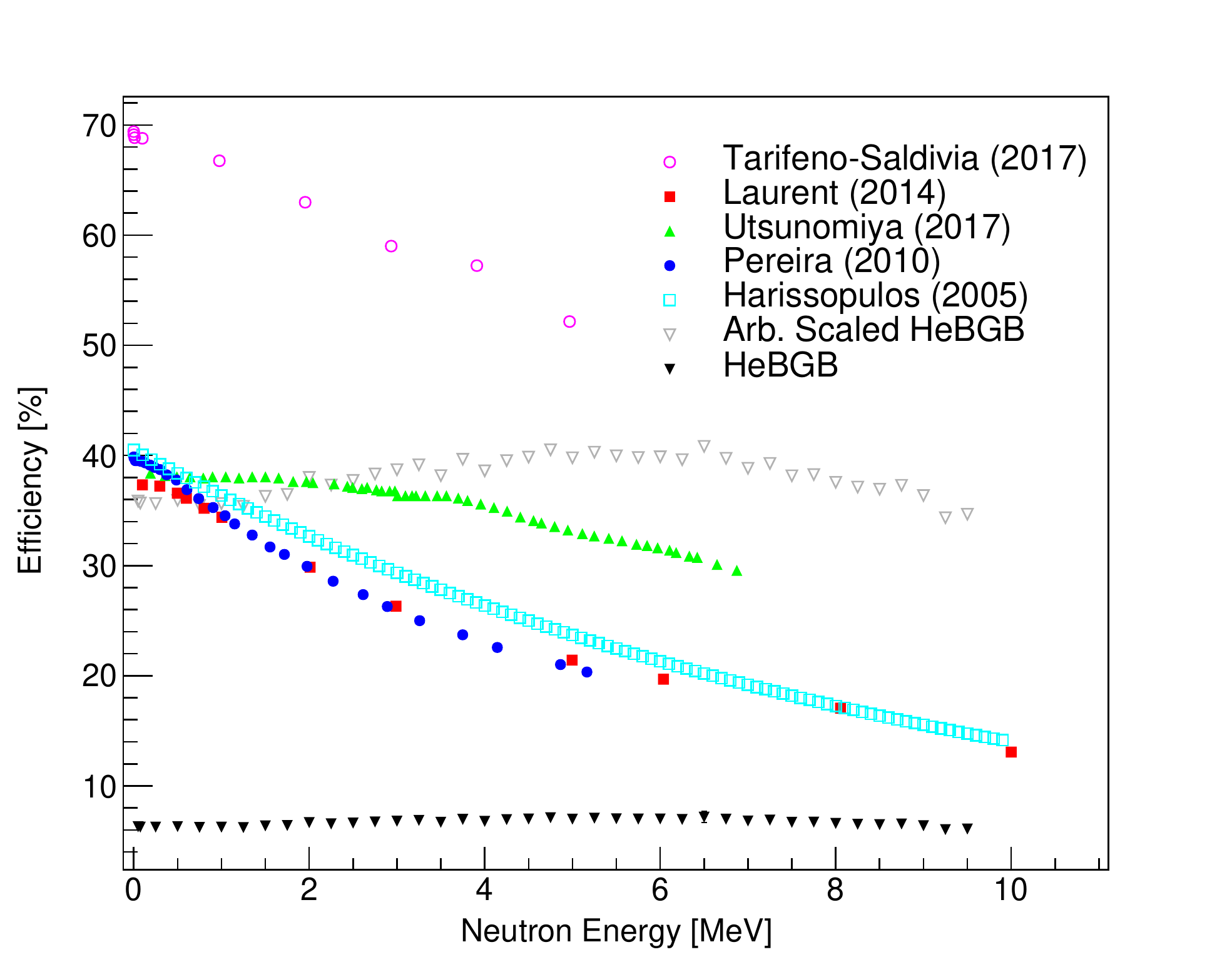}
        \caption{HeBGB efficiency (closed black triangles) calculated with MCNP for the final configuration as compared to the efficiency calculated for similar detectors~\cite[e.g.][]{Utsu17,Laur14,Tari17,Pere10,Hari05}. We also show an arbitrarily scaled HeBGB efficiency (open grey triangles)  for ease of comparison with the other detectors. Note that the final efficiency of HeBGB involves scaling the efficiency by ring and is shown in Figure~\ref{EffandData}. 
        \label{EffandE}}
        \end{center}
        \end{figure}

The optimum configuration for HeBGB was identified by taking the configuration with the smallest $\delta\epsilon$ from the initial configuration phase space and then taking smaller step sizes, 0.1~cm, in radius. Within these, the optimum configuration was defined as the one with the highest $\epsilon_{\rm avg}$. The geometry of the optimum configuration is shown in Figure~\ref{HeBGBView} with one outer ring of $^{3}{\rm He}$ detectors and two inner rings of BF$_{3}$ detectors. The corresponding neutron detection efficiency is shown in Figure~\ref{EffandE}. 

\begin{figure}[ht]
        \begin{center}
        \includegraphics[width=.9\columnwidth,angle=0]{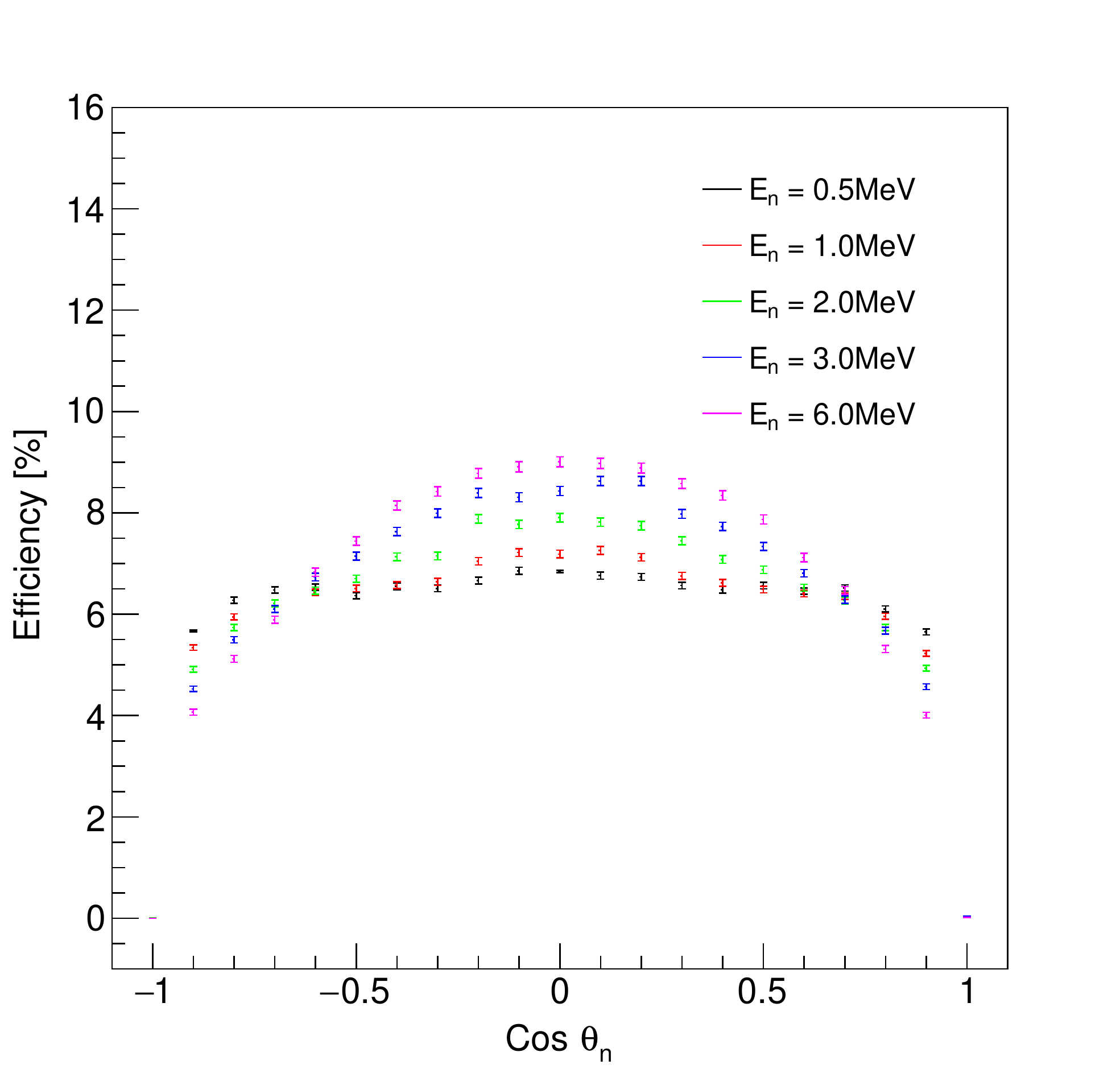}
        \caption{(color online) Calculated angular dependence of the HeBGB neutron detection efficiency for several initial neutron energies.
        \label{EffandCosTheta}}
        \end{center}
        \end{figure}

The effects of non isotropic distributions of neutrons was not considered during the optimization process but was explored after the optimum configuration was identified. Figure~\ref{EffandCosTheta} shows the efficiency as a function of angle for a select range of neutron energies with the final detector design configuration. The effects of anisotropy are greatest for higher energy neutrons, as high energy neutrons require more collisions within the moderator to thermalize and are therefore more likely to encounter the empty spaces left by the target ladder and beam pipe.
        
\section{Design Validation} 
\label{section:Experimental Validations}
\subsection{Experimental Setup} 

HeBGB is located at the Edwards Accelerator Laboratory at Ohio University~\cite{Meis17}. The experimental setup of the HeBGB detector system consists of a 0.5 cm gold plated collimator 48 cm upstream from the center of the detector, a 2 cm by 6 cm faraday cup located 37 cm downstream from the center of the detector, and a gold plated target ladder drive located in the center of the detector. The beam current is read separately from the collimator, cup, and target during the tuning process to ensure all beam is on target and is integrated during experiments to mitigate double-counting of accumulated charge due to escaping electrons.

\begin{figure}[ht]
        \begin{center}
        \includegraphics[width=.9\columnwidth,angle=0]{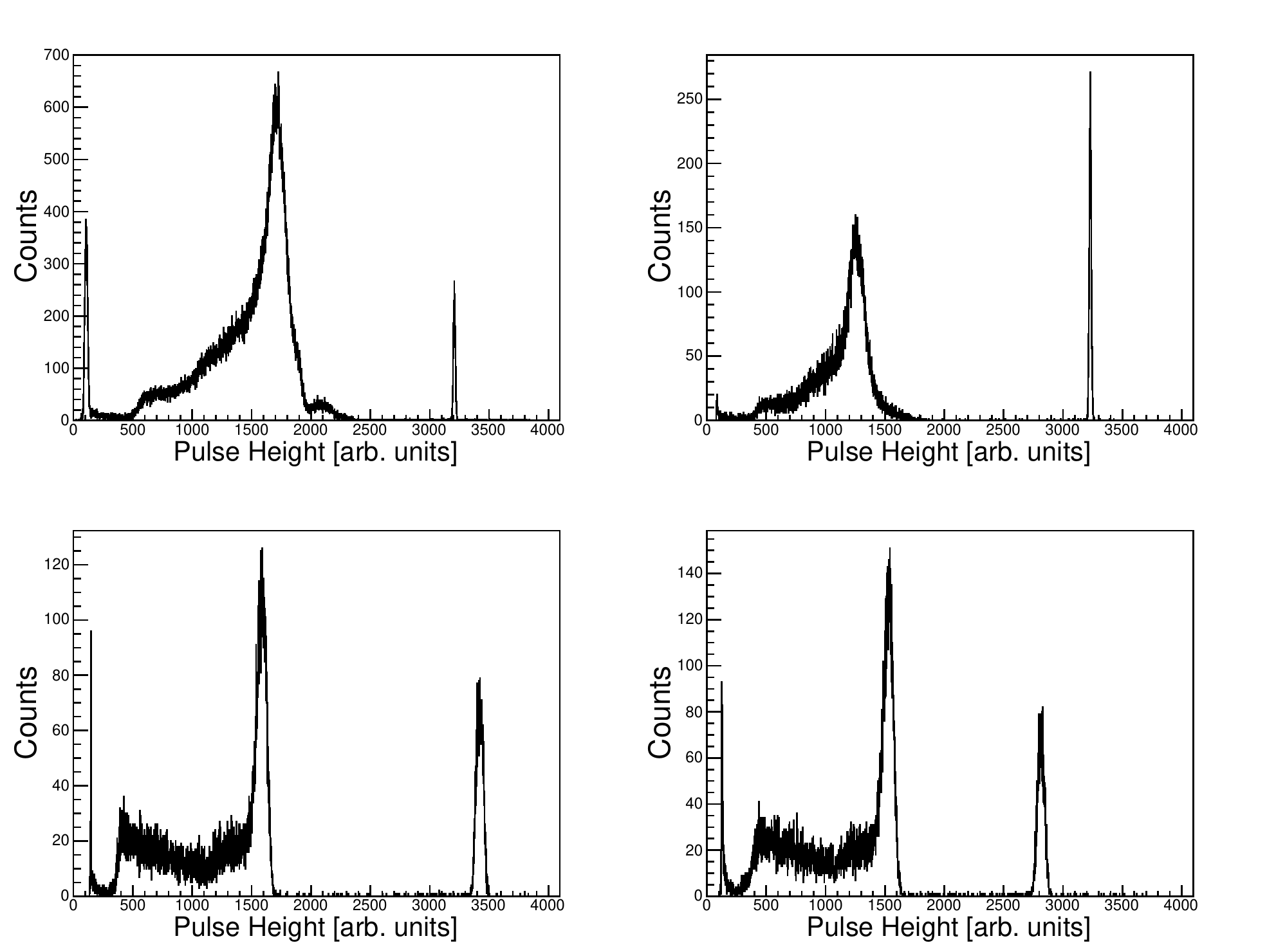}
        \caption{Sample spectra taken with a $^{252}$Cf source for the inner and middle ring BF$_{3}$ proportional counters (top row) and  segmented $^{3}$He outer ring (bottom row). The right most peak in each spectrum is the pulser, which is added to each detector set to measure deadtime. 
        \label{Spectra}}
        \end{center}
        \end{figure}

Neutron signals from each proportional counter are read together for both the inner and middle ring. The outer ring of $^{3}$He detectors is split into two sections consisting of eight counters in both the left and right hemispheres. 
Signals from the BF$_{3}$ and $^{3}$He detector sets are amplified using an EG\&G Ortec Model 142B and 142IH preamplifier, respectively. A single Ortec Model 419 Precision Pulse Generator 60 Hz signal is used in each detector set to measure deadtime, where the voltage is set sufficiently high to not interfere with the neutron counts in the spectra. Example spectra for each detector set are shown in Figure~\ref{Spectra}. 
        
\subsection{Validation Measurements}

The MCNP6 simulations of HeBGB were validated with three measurements. The first involved measuring neutrons from a calibrated $^{252}{\rm Cf}$ source. The source, provided by Eckert and Ziegler Isotope Products, contained 5.198 $\mu$Ci $^{252}$Cf referenced on 2017-10-01 with with quoted relative uncertainty of 3.6$\%$ ($@$99$\%$ confidence level). Mass and activity percentages of other Cf isotopes are quoted by the manufacturer and their contributing influences are accounted for in the calculation of activity at the measurement time. $^{248}$Cm decay products were last separated from this source in 2014 and are therefore not included in the calculation of neutron count rate, as the contribution is negligible~\cite{Rado14}. The source was placed at the center of HeBGB, in the usual target location, using a custom jig. Our efficiency uncertainty for this data point is due to the source activity uncertainty. The measured efficiency is compared with MCNP6 simulations of a $^{252}$Cf energy distribution using a Watt fission spectrum
\begin{equation}
N(E)=e^{-E_{n}/a}\sinh\left(\sqrt{bE_{n}}\right)
\label{eqn:watt}
\end{equation}
with parameters $a=1.18$ and $b=1.03419$~~\cite{Rado14}. These parameters are given in the MCNP6 documentation and are extremely close to the values given in the evaluation of Frohner ~\cite{Froh90}.  Within this distribution, the most probable energy is 0.70 MeV and the average is 2.13 MeV. When comparing to the efficiency simulated for monoenergetic neutrons, we set the energy at the most probable neutron energy and set the upper and lower error bars to encompass 68\% of the neutron spectrum about the mean. 


The second validation measurement was $^{51}{\rm V}(p,n)^{51}{\rm Cr}$ activation using 1.808 MeV protons. This results in neutrons with $E_{n}= (0.209-0.259$)~MeV. Following a 2.3~hr long irradiation, the activity of the irradiated sample was determined by counting 321~keV $\gamma$-rays emitted from $^{51}{\rm Cr}$ using a high-purity germanium detector (HPGe) located 10~cm from the sample, with both housed inside stacked lead bricks. The HPGe detector efficiency was determined using $^{152}{\rm Eu}$, $^{133}{\rm Ba}$, and $^{60}{\rm Co}$ sources, where a summing correction was performed following the method of Ref.~\cite{Semk90}. Via the activation equation, the number of neutrons emitted over the irradiation time is

\begin{equation}
    N=\frac{A(t)}{1-\exp\left(-{\rm ln}(2)t/t_{1/2}\right)},
    \label{eqn:activation}
\end{equation}
where $A(t)$ is target activity measured at time $t$ and the $^{51}{\rm Cr}$ half-life $t_{1/2}$ is from Ref.~\cite{Wang17}.
Our efficiency uncertainty for this point is primarily due to the 3.0\% HPGe detector efficiency uncertainty, which results from the fit of three standard $\gamma$-ray sources whose activity uncertainty are each $\sim3.1\%$. The measured efficiency of HeBGB is compared with MCNP6 simulations of this reaction assuming an isotropic distribution of neutrons in the center-of-mass frame, which follows from the statistical nature of this reaction.

\begin{figure}[ht]
        \begin{center}
        \includegraphics[width=.9\columnwidth,angle=0]{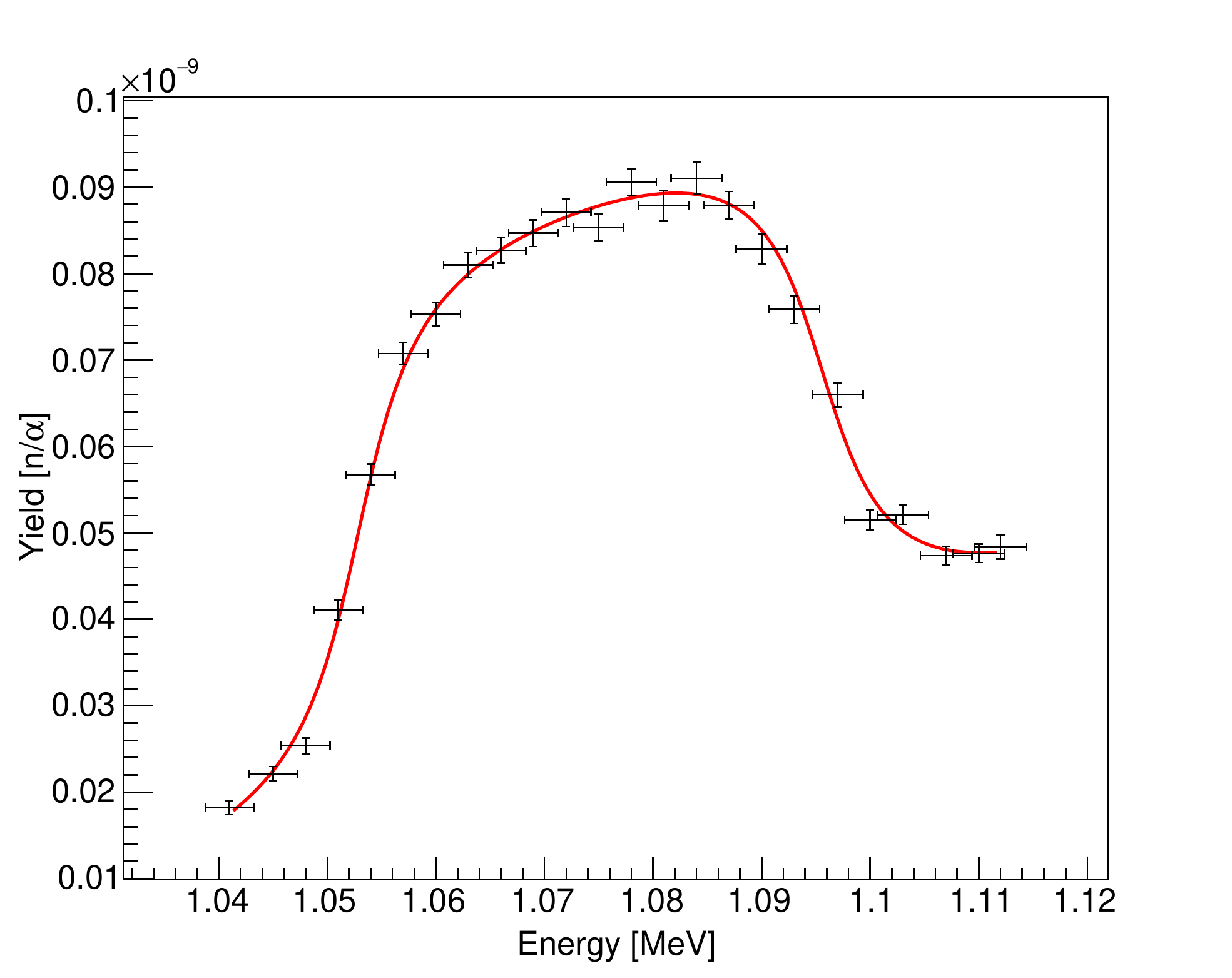}
        \caption{Neutron yield from $^{13}{\rm C}(\alpha,n)$, measured with HeBGB (points) and fit with a resonance yield curve and a linear non-resonant background contribution (after ambient background subtraction).
        \label{C13Res}}
        \end{center}
        \end{figure}

The third validation measurement was the measurement of the well-characterized resonance in $^{13}{\rm C}(\alpha,n)$ at laboratory $\alpha$ energy $E_{\alpha}=1.053$~MeV, resulting in $E_{n}\sim(2.45-3.27$)~MeV. The neutron yield was measured for $E_{\alpha}=(1.041-1.112$)~MeV, as shown in Figure~\ref{C13Res}. Background measurements with an empty target frame were performed at each energy and were found to be at the level of room background. Room background was measured before, after and during the scan. An average room background contribution was calculated and subtracted from the data runs. The data was fit with a standard resonant yield curve~\cite{Iliadis} combined with a linear background intended to account for the non-resonant contribution to the reaction:
\begin{equation}
   Y(E_{\alpha})=a+bE_{\alpha}+Y^{\rm meas}_{\rm max}\left[\arctan\left(\frac{E_{\alpha}-E_{\rm r}}{\Gamma/2 + c }\right)-\left(\frac{E_{\alpha}-E_{\rm r}-\Delta E}{\Gamma/2 + d}\right)\right].
    \label{eqn:ResFit}
\end{equation}
Here, $a$ and $b$ are terms describing the residual background, $c$ and $d$ are diffuseness parameters describing the beam resolution and straggling,  $Y^{\rm meas}_{\rm max}$ is the maximum yield in the resonance curve, $E_{\rm r}=1.053$~MeV is the resonance energy, $\Delta E$ is the total energy loss of the $\alpha$-particle within the $^{13}{\rm C}$ target, and $\Gamma$ is the resonance width, which is fixed to the value reported by Ref.~\cite{Bair73} during the fit. 
The predicted maximum yield is determined by the resonance strength $\omega\gamma$ and stopping power $\varepsilon_{\rm r}$ (each in the center-of-mass frame) of an $\alpha$-particle in the $^{13}{\rm C}$ target at $E_{\rm r}$~\cite{Iliadis}:
\begin{equation}
    Y^{\rm calc}_{\rm max}=\frac{\lambda_{r}^2\omega\gamma}{2\varepsilon_{\rm r}}.
    \label{eqn:Ymax}
\end{equation}
For $\omega\gamma$, we employ the results from a recent reanalysis of Refs.~\cite{Kell89,Brun92,Brun93} that uses an improved model for neutron detection efficiency and takes advantage of updated nuclear data. That reanalysis will be reported in an upcoming work~\cite{BruneTBD} . For $\varepsilon_{\rm r}$, we take a weighted average of the world data for $\alpha$-particles in $^{13}{\rm C}$ at this energy~\cite{Mont17}. Our detector efficiency is $Y^{\rm meas}_{\rm max}/Y^{\rm calc}_{\rm max}$. In order to to compare this efficiency to the results of Figure~\ref{EffandE}, we simulated the impact of the angular distribution as determined by $R$-matrix calculations, also described by Ref.~\cite{BruneTBD}. The 8.2\% uncertainty for this efficiency determination is primarily due to the uncertainties for $\varepsilon_{\rm r}$ (6.5\%) and $\omega\gamma$ (5.0\%).

\section{Efficiency Corrections} 
\label{section:Efficiency Corrections}

\begin{figure}[ht!]
        \begin{center}
        \includegraphics[width=.9\columnwidth,angle=0]{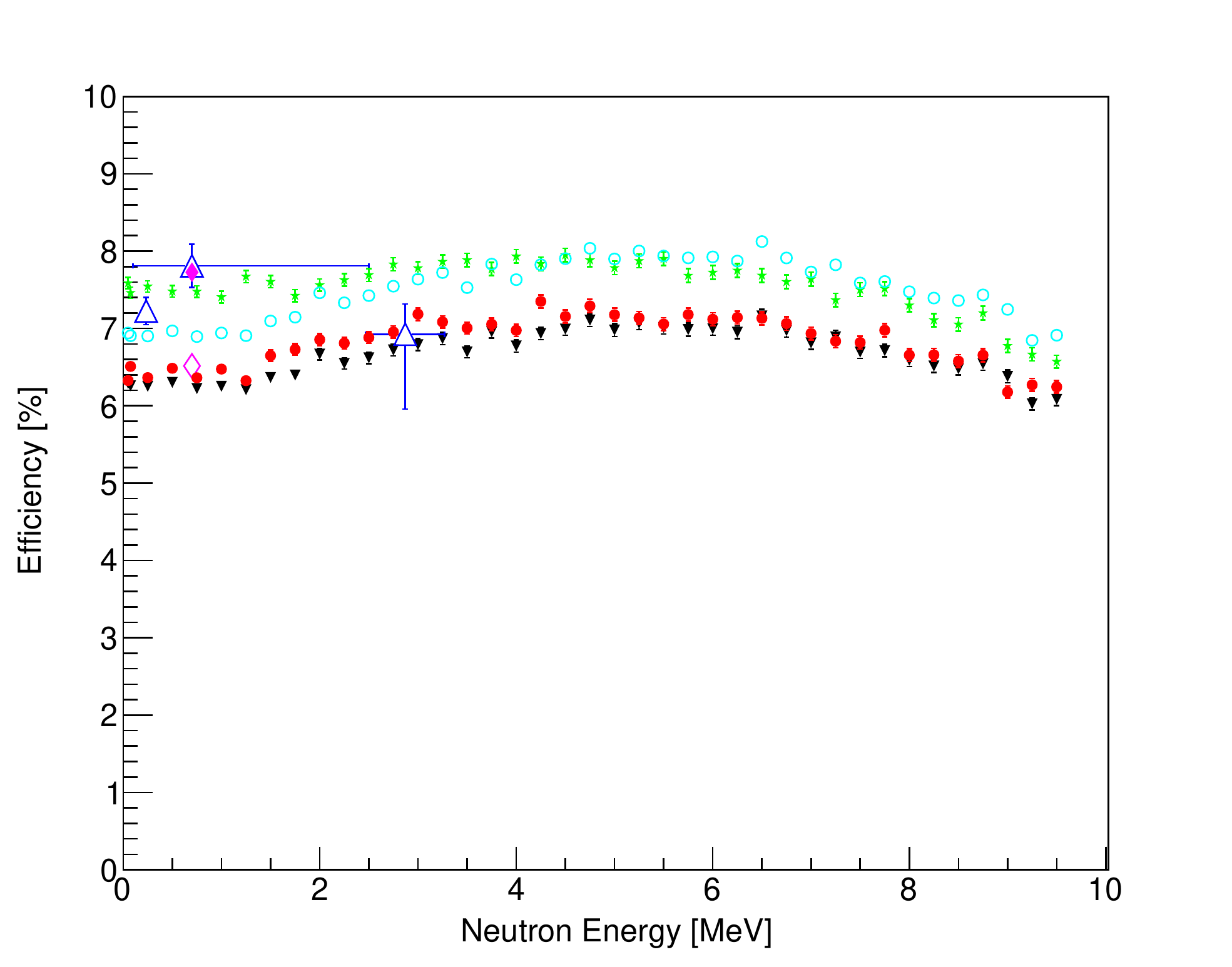}
        \caption{(color online) Measured neutron detection efficiency for HeBGB (open blue triangles), compared to various MCNP calculations. The black filled triangles are the original MCNP simulations, red circles are the simulations with modified polyethylene density, green stars are simulations with modified BF$_{3}$ pressure, and the open cycan circles are the scaled simulation results to data. Simulations of $^{252}$Cf are shown in pink with an open diamond for the original simulation with no modifications and close diamond for the simulation with modified BF$_{3}$ pressure.
        \label{EffandData}}
        \end{center}
        \end{figure}

The results of the validation measurements and MCNP6 simulations are compared in Figure~\ref{EffandData}. The measured efficiencies are larger than simulation results. To account for this discrepancy, we have examined the effects of modifying both the polyethylene density and gas pressure in the MCNP calculations. A sample piece provided by the manufacturer from the same batch as the larger polyethylene pieces measured (0.95 $\pm$ 0.2) g\,cm$^{-3}$ and was used for the initial simulations. Ranges of density for ultrahigh molecular weight polyethylene are from (0.931 -- 0.949)~gcm$^{-3}$~\cite{Huss20}. Using densities within this range resulted in negligible differences between calculation results, consistent with the findings of earlier investigations~\cite{Char03}. An increased BF$_{3}$ pressure from 550 Torr to 760 Torr was also simulated and found to more closely resemble the data; however, the manufacturer quoted uncertainty in proportional counter fill pressure is only 0.25\%\footnote{This fill-pressure uncertainty was communicated by Baker-Hughes, which is the modern-day parent company of the manufacturer of our BF$_{3}$ detectors, Reuter-Stokes. We have no original records from the detectors and all specifications come from interpreting the detector model numbers.}. Alternatively, keeping the original fill pressure and scaling each ring by a scale factor found by minimizing $\chi^{2}$ between the measurements and calculated results for each ring, as performed in Refs.~\cite{Pere10,Csed21}, results in a similar efficiency to the adjusted BF$_{3}$ pressure. Scale factors 1.104, 1.177, and 1.159 are used for the inner ring, middle ring, and outer ring, respectively, to arrive at the final detector efficiency. An additional error on the final efficiency is assigned by taking the average difference between the three measured efficiencies and the simulated efficiencies evaluated at the neutron energy of each measurement. This results in an additional error of 0.6\% to be added to the error from the spread in efficiency over the neutron energy range using Equation~\ref{eqn:EfEr}. The error from the spread in efficiency is 0.63\%. The final scaled detector efficiency over the neutron energy range 0.01~MeV -- 9.00~MeV is ($7.5\pm 1.2$) \%.

\section{Average Initial Neutron Energy Determination} 
\label{section:Average Initial Neutron Energy Determination}
\begin{figure}[ht!]
        \begin{center}
        \includegraphics[width=.9\columnwidth,angle=0]{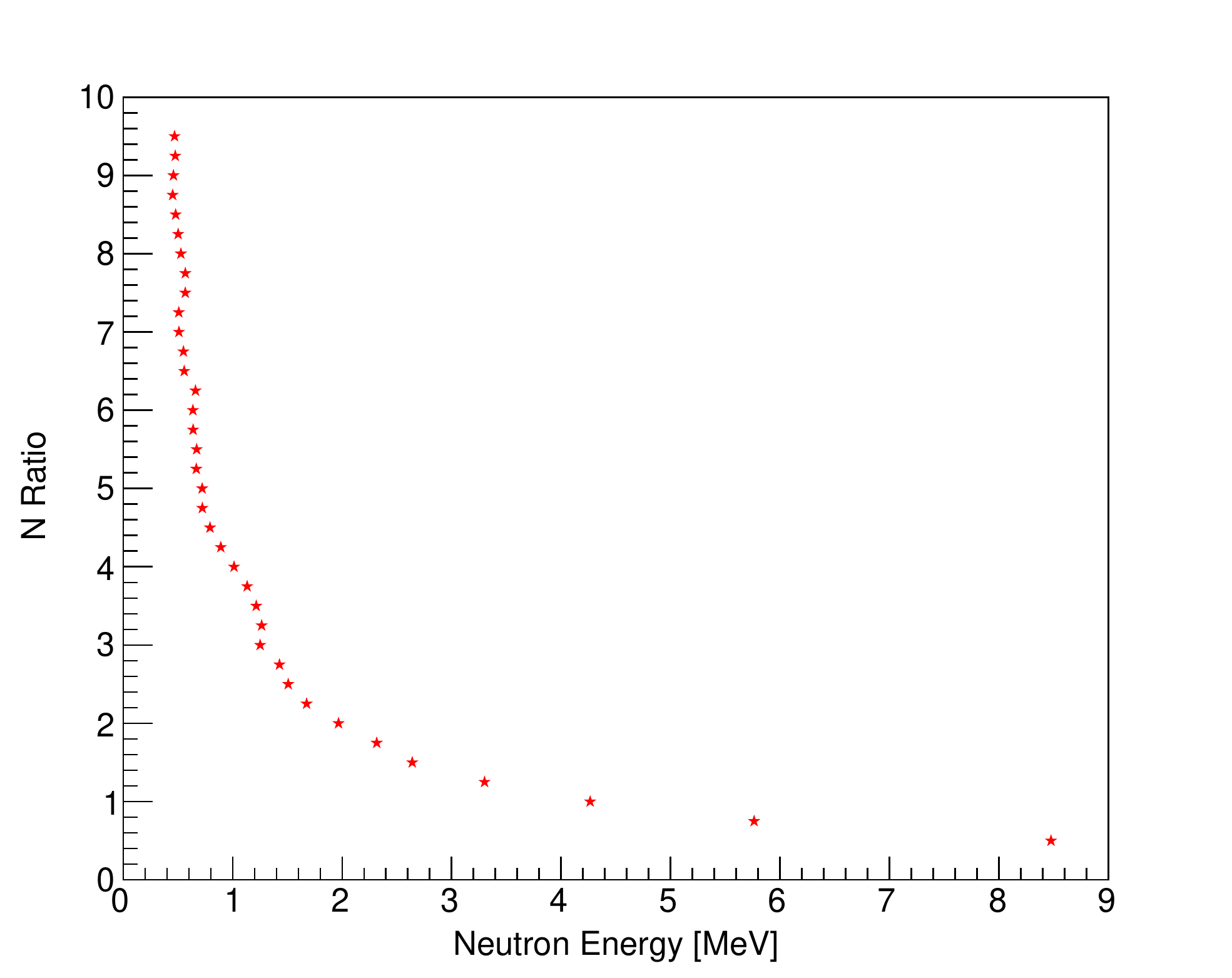}
        \caption{MCNP results for the ratio of counts detected within the inner ring of HeBGB to counts within the outer two rings for several initial neutron energies.
        \label{RingRatioSim}}
        \end{center}
        \end{figure}

Due to the statistical nature of the path of the neutrons in the moderator, the initial energy of a particular neutron emitted from the target location is unknown. However, average energies can be determined from the ratio of neutrons detected in the inner ring to neutrons detected in the combined two outer rings. MCNP results in Figure~\ref{RingRatioSim} show the trend in the ring ratio with increasing initial neutron energy, where larger energies result in a smaller ratio. Therefore, we can use the measured ring ratio to obtain a coarse estimate of the average initial neutron energy for a particular measurement.

\begin{figure}[ht!] 
        \begin{center}
        \includegraphics[width=.9\columnwidth,angle=0]{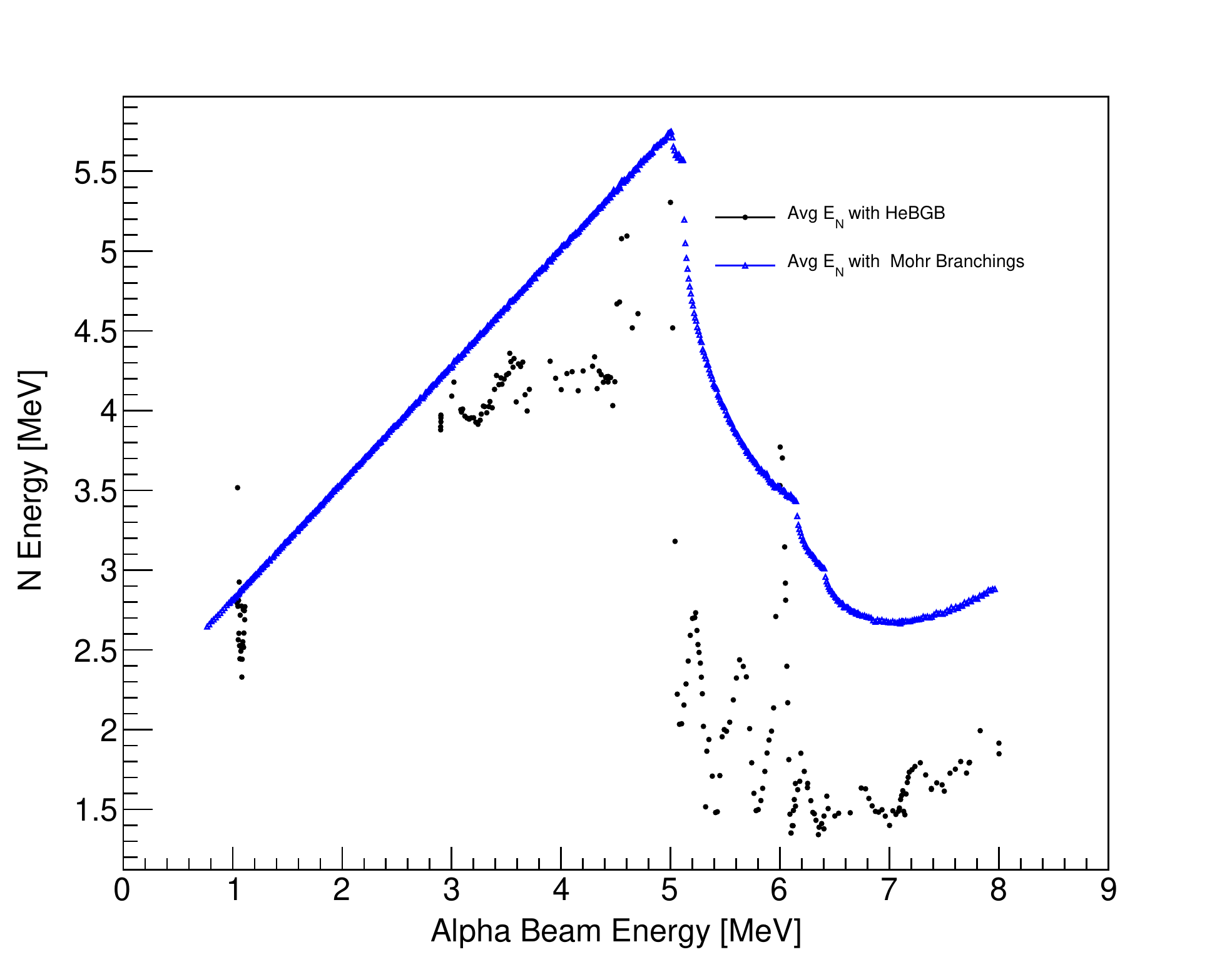}
        \caption{(color online) Average initial neutron energy from $^{13}{\rm C}(\alpha,n)$ determined using the ring ratio method (black solid points) compared to the average neutron energy predicted in Ref.~\cite{Mohr18}. 
        \label{C13NeutronEnergy}}
        \end{center}
        \end{figure}

To demonstrate this method, we have calculated the ring ratio of data from a measurement of $^{13}$C($\alpha$,n)  from E$_{\alpha}=(1.041-1.112)$~MeV and ($3.00-8.00$)~MeV. The full details of the measurement and results will be presented in a future publication. Only the pertinent details for the ring-ratio measurement are discussed here. The ratio of neutron detections within the inner ring to detections in the outer rings is calculated for each $^{13}{\rm C}(\alpha,n)$ measurement energy and that ratio is mapped on to the ring ratio calculations for an initial neutron energy shown in Figure~\ref{RingRatioSim}. Figure~\ref{C13NeutronEnergy} shows the resulting estimate for initial neutron energy over the measured $\alpha$ energy range together with the average neutron energy calculated from kinematics for all de-excitations to the $^{16}$O nucleus. The branching ratios for each state are taken from the Hauser-Feshbach based calculations of Ref.~\cite{Mohr18}. 

Figure~\ref{C13NeutronEnergy} shows that the ring ratio method qualitatively agrees with the predicted average neutron energy. The disagreement between $E_{\alpha} \sim $ (3-5)~MeV is likely due to the shallow slope of the ring-ratio versus neutron-energy trend shown in Figure~\ref{RingRatioSim}. Given the shallow slope, small changes in the ring ratio can lead to large shifts in the estimated initial neutron energy. Above $\sim$5~MeV, decay branchings open to excited states of $^{16}{\rm O}$ and it is possible that some of the disagreement is due to inaccurate branching ratio estimates in Ref~\cite{Mohr18}. Our preliminary investigations indicate that the large fluctuations in the estimated neutron energy are due to angular distribution impacts on the ring ratio.


\section{Conclusions} 
\label{section:Conclusions}
  
We developed the HeBGB neutron detector, designed to accurately count neutrons emitted from $(\alpha,n)$ reactions. Monte Carlo calculations were used to optimize neutron-sensitive proportional counter placement within a polyethylene moderator in order to achieve a near-constant neutron detection efficiency, ($7.5\pm 1.2$) \%, over the neutron energy range 0.01~MeV -- 9.00~MeV. Note, 0.6\% of this error is from the scaling of simulation to experimental data and 0.63\% is from the spread in efficiency over the neutron energy range, so the uncertainty could be reduced to $\pm$ 0.6\% if the neutron energy distribution were known or alternatively if an improved simulation would not require scaling to data. The detector performance was validated using neutron source and neutron-emitting reaction measurements. It was demonstrated that coarse constraints on the initial neutron energy can be obtained by assessing the ratio of neutron detections in the inner ring of HeBGB detectors to detections in the outer ring of HeBGB detectors. The HeBGB detector will enable accurate measurements of $(\alpha,n)$ cross sections of interest for astrophysics and applications, providing a near-constant neutron detection efficiency, avoiding a significant systematic uncertainty present in earlier set-ups.

\paragraph{}
\bibliographystyle{JHEP}
\bibliography{Main}


\end{document}